\documentclass{article}
\usepackage{amsmath,amsfonts,amssymb,amsthm,bbm,booktabs, setspace}
\usepackage{algorithm} 
\usepackage{algpseudocode} 
\usepackage{mathtools}
\usepackage{commath}
\usepackage{graphicx}
\usepackage{xcolor}
\usepackage{setspace}
\usepackage{hyperref}
\usepackage[margin = 1 in]{geometry}
\newcommand\mc{\mathcal}

\newtheorem{definition}{Definition}[subsection]
\newtheorem{theorem}{Theorem}[subsection]
\newtheorem{corollary}{Corollary}[subsection]
\newtheorem{lemma}{Lemma}[subsection]
\newtheorem{example}{Example}[subsection]
\newtheorem{remark}{Remark}[subsection]
\newtheorem{proposition}{Proposition}[subsection]

\newcommand{\supp}{\mathrm{supp}}
\newcommand{\sgn}{\mathrm{sign}}
\newcommand{\nz}{\mathrm{nz}}

\title{Support Recovery in One-bit Compressed Sensing with Near-Optimal Measurements and Sublinear Time}
\author{Xiaxin Li \and Arya Mazumdar
\thanks{The authors are with the Hal{\i}c{\i}o\u{g}lu Data Science Institute, University of California, San Diego. emails: \texttt{\{xil095, arya\}@ucsd.edu}.
}
\thanks{This work was supported in part by NSF awards 2217058 and 2112665.}
}
\date{}
 
\begin{document}
\maketitle

\begin{abstract}
    One-bit compressed sensing (1bCS) addresses the recovery of sparse signals from highly quantized measurements, retaining only the sign of each linear measurement. From a data compression perspective, the one-bit measurements form a compact binary representation of sparse signals. The support recovery problem seeks to recover the support of an unknown signal $x\in\mathbb{R}^n$, $\supp(x)$, from $y=\sgn(Ax)$, where $A\in\mathbb{R}^{m\times n}$ is the measurement matrix and $|\supp(x)|\le k\ll n$. Existing methods seek to minimize the number of measurements but often incur $\Omega(n)$ decoding complexity, limiting their applicability to large-scale problems.

We propose new 1bCS schemes that achieve sublinear decoding complexity while maintaining near-optimal measurement bounds. For universal support recovery, our framework provides: (i) exact recovery with $m=O(k^2\log(n/k)\log n)$ measurements and decoding complexity $D=O(km)$, and (ii) $\epsilon$-approximate recovery with $m=O(k\epsilon^{-1}\log(n/k)\log n)$ and $D=O(\epsilon^{-1}m)$. For probabilistic exact recovery, we design a scheme with $m=O(k\log k\log n)$ and $D=O(m)$, achieving vanishing error probability. Our schemes leverage ideas from group testing to achieve near-optimal support compression with substantially reduced decoding complexity.

\end{abstract}

\section{Introduction}\label{sec:intro}
Compressed sensing~\cite{CandesT06,Donoho06} aims to recover a high-dimensional sparse signal from a small number of measurements and has found applications in signal processing, wireless communications, and machine learning. Despite its success, classical compressed sensing typically requires high-precision measurements, leading to substantial computational and storage costs. To address this limitation, a quantized version known as one-bit compressed sensing (1bCS) was introduced~\cite{boufounos20081} and has since been extensively studied, 
e.g. in~\cite{li2018survey,plan2013one,baraniuk2010model,acharya2017improved,gopi2013one,zhang2014efficient,nakos2017fast,matsumoto2024robust,matsumoto2024binary}. In 1bCS, the goal is to recover an unknown signal $x\in\mathbb{R}^n$ from measurements of the form $y = \sgn(Ax)$, where $A \in \mathbb{R}^{m \times n}$ is the measurement matrix, and $\sgn(\cdot)$ is applied coordinate-wise, mapping non-negative entries to $1$ and negative entries to $-1$. It is assumed that $x$ is $k$-sparse, i.e., $|\supp(x)| \le k \ll n$.

The 1bCS problem is typically studied under two settings: support recovery and approximate vector recovery. Support recovery aims to recover the nonzero coordinates of $x$, i.e., $\supp(x)$, while approximate vector recovery seeks a vector $x'$ close to $x$ under a given norm. These two settings are closely related: once the support of $x$ is known, an approximate recovery of $x$ can often be efficiently computed ~\cite{gopi2013one,flodin2019superset}. \emph{In this paper, we focus exclusively on support recovery.}

One-bit compressed sensing, and any such sparse recovery problems, bear a close connection to data compression from an information-theoretic perspective. The measurement process maps a high-dimensional sparse signal to a significantly shorter binary representation, with each measurement contributing only one bit of information. When the recovery objective is restricted to support recovery, this process can be viewed as compressing the unknown support into a binary codeword from which the support is subsequently decoded. Since there are about $\binom{n}{k}$ possible supports, about
$
\log_2 \binom{n}{k}
=
\Theta\!\left(k\log\frac{n}{k}\right)
$
bits are required to distinguish them. From this viewpoint, a one-bit compressed sensing scheme can be regarded as an encoder--decoder pair for sparse support information, where the sensing matrix performs the encoding and the recovery algorithm performs the decoding. Consequently, the effectiveness of such a compression framework depends not only on approaching the information-theoretic minimum number of one-bit measurements, but also on recovering the desired support with low computational complexity. Indeed, a compression scheme is practically valuable only if its compressed representation can be decoded efficiently. Our proposed schemes address both objectives by employing a nearly order-optimal number of one-bit measurements while significantly reducing the decoding complexity compared with existing approaches, thereby providing an efficient compression-and-decoding framework for sparse support information.

Specifically, most existing 1bCS support recovery schemes use decoding algorithms that individually check for each of the $n$ coordinates whether it is part of the unknown support. 
This requires an $\Omega(n)$ runtime, e.g. \cite{gopi2013one,acharya2017improved,matsumoto2023improved}. However, the sparsity of $x$ and the structural properties of certain measurement matrices allow for sublinear-time decoding. A closely related problem is {\em nonadaptive group testing}, where the goal is to recover a $k$-sparse binary vector $x \in \{0,1\}^n$ from $y = A \circ x$, with $A \in \{0,1\}^{m \times n}$ and ``$\circ$'' denoting matrix multiplication with OR and AND replacing addition and multiplication. Structural properties in group testing have been leveraged to achieve highly efficient sublinear decoding algorithms.

Motivated by these connections, our aim is to design 1bCS support recovery algorithms with sublinear decoding complexity, ideally $O(m)$, with only additional factors involving $k$, $\log k$, and $\log n$. To achieve this, we adapt techniques from group testing, where efficient decoding algorithms have been extensively developed~\cite{lee2019saffron,cheraghchi2020combinatorial,cai2017efficient,guruswami2023noise}.

Next, we introduce the notations used throughout the paper and formally define the support recovery problem.

\subsection{Basic Notations}
For $n \in \mathbb{N}^+$, let $[n] := \{1,2,\dots,n\}$. For $r \in \mathbb{R}$, let $\lceil r \rceil$ denote the smallest integer greater than or equal to $r$. For a vector $v = [v_1, v_2, \dots, v_n] \in \mathbb{R}^n$ (or $\{0,1\}^n$), define its support as
\(
\supp(v) := \{ i \in [n] : v_i \neq 0 \}.
\)
For a finite set $S$, let $|S|$ denote its cardinality. Let $0^n$ denote the all-zero vector in $\mathbb{R}^n$.

For a matrix $A \in \mathbb{R}^{m \times n}$ (or $\{0,1\}^{m \times n}$), denote its $i$-th row and $j$-th column by $A^i$ and $A_j$, respectively. The $(i,j)$-th entry of $A$ is denoted by $A_{ij}$; when convenient, we may also write $A^i[j]$ or $A_j[i]$. For index sets $S = \{s_1, \dots, s_u\} \subseteq [m]$ and $T = \{t_1, \dots, t_v\} \subseteq [n]$, listed in increasing order, define the submatrices
\(
A^{S} := [A^{s_1}; A^{s_2}; \dots; A^{s_u}],
A_T := [A_{t_1}, A_{t_2}, \dots, A_{t_v}],
\)
corresponding to the rows indexed by $S$ and the columns indexed by $T$, respectively.

\subsection{Support Recovery in One-Bit Compressed Sensing}\label{sec:1bCS}

The \emph{sign function} $\sgn : \mathbb{R} \to \{-1,1\}$ is defined as
\[
\sgn(a) =
\begin{cases}
1 & \text{if } a \geq 0,\\
-1 & \text{if } a < 0.
\end{cases}
\]
For vectors and matrices, $\sgn(\cdot)$ is applied entrywise.

In the support recovery problem for one-bit compressed sensing (1bCS), the goal is to recover the support of a $k$-sparse \emph{signal vector} $x \in \mathbb{R}^n$, i.e. $|\supp(x)| \leq k$, from measurements
\(
y := \sgn(Ax),
\)
where $A \in \mathbb{R}^{m \times n}$ is the \emph{sensing matrix} and $y \in \{-1,1\}^m$ is the \emph{result vector}. The objective is to design $A$ with as few measurements $m$ as possible while enabling efficient decoding. In this work, we primarily focus on minimizing the decoding complexity, allowing the number of measurements to be slightly suboptimal.

For convenience, we define the \emph{nonzero function} $\nz : \mathbb{R} \to \{0,1\}$ by
\[
\nz(a) =
\begin{cases}
1 & \text{if } a \neq 0,\\
0 & \text{if } a = 0.
\end{cases}
\]
Note that 
\begin{equation}\label{eqn:nz}
    \nz(a) = \frac12(1-\sgn(a)\cdot\sgn(-a)).
\end{equation} As before, $\nz(\cdot)$ is applied entry-wise to vectors and matrices.

\textbf{Throughout the paper,} (i) we use the notation $n, m, k, x, y, A$ as defined above, (ii) adopt $\nz(\cdot)$ instead of $\sgn(\cdot)$ as measurements, since $\nz(\cdot)$ can be reproduced from $\sgn(\cdot)$ by \eqref{eqn:nz} above, not affecting the asymptotic scaling  of number of measurements, and (iii) use the Word-RAM model for analyzing computational complexities. 

We distinguish between two types of recovery guarantees for a given sensing matrix~\cite{acharya2017improved}. A \emph{for-all} (or \emph{universal}) scheme succeeds simultaneously for all feasible signal vectors $x$. In contrast, a \emph{for-each} (or \emph{probabilistic}) scheme succeeds for most signal vectors, while failure is restricted to a vanishing fraction of them.
To be precise, in this later setting we can assume that the $k$-sparse support of the signal vector is chosen uniformly at random from all possibilities, and then show that a scheme succeeds with high probability (with respect to the randomness in the scheme and the signal). Alternatively, we can take a fixed signal and further randomize the scheme by first permuting the coordinates uniformly. For randomized constructions, the for-all schemes guarantee uniform success over all $x$ with high probability, while the for-each schemes guarantee success for each fixed $x$ individually.

Let $S \subseteq [n]$ denote the output of a recovery algorithm. We consider two notions of support recovery:

\begin{itemize}
    \item \emph{Exact support recovery:} \[S = \supp(x).\]
    \item \emph{$\epsilon$-approximate support recovery:} 
    \begin{equation}\label{eqn:asr}
        \begin{gathered}
            |S \setminus \supp(x)| \leq \epsilon |\supp(x)| 
            \text{, and } 
            |\supp(x) \setminus S| \leq \epsilon |\supp(x)|.
        \end{gathered}
    \end{equation}
\end{itemize}
The latter notion allows for a small fraction of both false positives and false negatives~\cite{matsumoto2023improved}.

To the best of our knowledge, very few prior works achieve sublinear-time decoding for 1bCS support recovery.  A very recent result is~\cite{yang2025sublinear}, giving a probabilistic exact recovery scheme with 
$m = O\big(k^2 \log^2 (n) (\log^2 k + (\log\log n)^2)\big)$
measurements and decoding time $O(m)$. 
Another notable result is \cite{nakos2017fast}, giving a universal exact recovery scheme with decoding time $O(k^3\mathrm{poly}(\log n))$, but the number of measurements is $O(k^3\log n)$, which is suboptimal by a factor of $\Omega(k)$.

In this paper, we develop sublinear-time algorithms for:
\begin{itemize} 
    \item Universal $\epsilon$-approximate recovery,
    \item Universal exact recovery, and
    \item Probabilistic exact recovery.
\end{itemize}

We collectively refer to these schemes as \textsc{EDOCS} (Efficient Decoding One-bit Compressed Sensing). In Table~\ref{tab:comparison} below, we summarize all our main results and compare them with the most recent prior works and lower bounds on the number of measurements.  

\begin{table}[H]
\centering
\setlength{\tabcolsep}{2pt}
\small
\begin{tabular}{@{}lcccc@{}}
\toprule

\multicolumn{5}{c}{\textbf{Universal $\epsilon$-approximate recovery}}\\
\midrule
& Lower bound \cite{matsumoto2023improved}
& \multicolumn{2}{c}{\cite{matsumoto2023improved}}
& EDOCS-AA (Theorem~\ref{thm:aa})\\
\midrule
\# Measurements
& $\Omega\left(k\epsilon^{-1}\dfrac{\log(\frac{n}{k\epsilon})}{\log(\frac{k}{\epsilon})}\right)$
& \multicolumn{2}{c}{$O(k\epsilon^{-1}\log(\frac{n}{k}))$}
& $O(k\epsilon^{-1}\log(\frac{n}{k})\log n)$\\
Decoding complexity
& N/A
& \multicolumn{2}{c}{$O(n\epsilon^{-1}\log(\frac{n}{k}))$}
& $O(k\epsilon^{-2}\log(\frac{n}{k})\log n)$\\

\midrule
\multicolumn{5}{c}{\textbf{Universal exact recovery}}\\
\midrule
& Lower bound \cite{acharya2017improved}
& \cite{acharya2017improved}
& \cite{nakos2017fast}
& EDOCS-AE (Theorem~\ref{thm:ae})\\
\midrule
\# Measurements
& $\Omega\left(k^2\dfrac{\log n}{\log k}\right)$
& $O(k^2\log n)$
& $O(k^3\log n)$
& $O(k^2\log(\frac{n}{k})\log n)$\\
Decoding complexity
& N/A
& $O(nk\log n)$
& $O(k^3\mathrm{poly}(\log n))$
& $O(k^3\log(\frac{n}{k})\log n)$\\

\midrule
\multicolumn{5}{c}{\textbf{Probabilistic exact recovery}}\\
\midrule
& Lower bound (Folklore)
& \multicolumn{2}{c}{\cite{yang2025sublinear}}
& EDOCS-EE (Theorem~\ref{thm:ee})\\
\midrule
\# Measurements
& $\Omega\left(k\log \frac{n}{k}\right)$
& \multicolumn{2}{c}{$O(k^2\log^2 n\,(\log^2 k+(\log\log n)^2))$}
& $O(k\log k\log n)$\\
Decoding complexity
& N/A
& \multicolumn{2}{c}{$O(k^2\log^2 n\,(\log^2 k+(\log\log n)^2))$}
& $O(k\log k\log n)$\\

\bottomrule
\end{tabular}

\vspace{4pt}

\caption{Our results under different recovery guarantees, with comparison to the most recent prior works.}
\label{tab:comparison}
\end{table}

\paragraph{Organization.} The rest of the paper is organized as follows. Section \ref{sec:USR} presents our universal exact-recovery and universal $\epsilon$-approximate recovery schemes (see, Theorem~\ref{thm:aa},~\ref{thm:ae}). Section \ref{sec:RSR} presents our probabilistic exact-recovery scheme (see, Theorem~\ref{thm:ee}). Section \ref{sec:sum} summarizes our results. 

\section{EDOCS: Universal support recovery}\label{sec:USR}
In this section, we present the \textsc{EDOCS} schemes for universal support recovery, which address both $\epsilon$-approximate and exact recovery. We denote these schemes by \textsc{EDOCS-AA} (\emph{for-all, approximate}) and \textsc{EDOCS-AE} (\emph{for-all, exact}), respectively. The two schemes share the same overall framework and differ only in the choice of a key combinatorial matrix, as will be specified later.

As discussed in the introduction, most existing decoding algorithms have a decoding complexity of $\Omega(n)$. Our approach avoids this bottleneck by first identifying a small candidate set that approximates the support and then restricting further computation to this set.

More concretely, in \textsc{EDOCS-AA}, we first construct a superset $S \subseteq [n]$ such that
\[
|S \cap \supp(x)| > (1-\epsilon)|\supp(x)|.
\]
We then refine $S$ by examining only its elements and removing false positives, so that
\[
|S \setminus \supp(x)| < \epsilon |\supp(x)|, \text{ and }
|\supp(x) \setminus S| < \epsilon |\supp(x)|.
\]
thus achieving $\epsilon$-approximate support recovery.

Similarly, in \textsc{EDOCS-AE}, we first construct a superset $S \supseteq \supp(x)$, and then prune $S$ so that $S = \supp(x)$, achieving exact recovery.

In both schemes, the construction of the candidate set $S$ relies on carefully designed combinatorial matrices, which we introduce next.

\subsection{Combinatorial matrices} 
Our constructions rely on matrices with specific combinatorial properties. In particular, we use the notion of $(n,m,d,k,l,\alpha)$-list union-free families and their corresponding matrices, introduced in~\cite{matsumoto2023improved}. We recall these definitions below.

\begin{definition}\label{def:basic}
(\cite[Definition~6]{matsumoto2023improved})
Let $0 \leq \alpha \leq 1$. A family of sets $\mc{F} = \{B_1, B_2, \dots, B_n\}$ with $B_i \subseteq [m]$ and $|B_i| = d$ for all $i \in [n]$ is called a $(n,m,d,k,l,\alpha)$-\emph{list union-free (UF) family} if for any pair of disjoint sets $S, T \subseteq [n]$ with $|S| = l$ and $|T| = k$, there exists an index $j \in S$ such that
\[
\left| B_j \cap \bigcup_{i \in (T \cup S)\setminus \{j\}} B_i \right| < \alpha d.
\]

Given such a family $\mc{F}$, the associated binary matrix $M \in \{0,1\}^{m \times n}$ is defined by
\[
M_{ij} =
\begin{cases}
1 & \text{if } i \in B_j,\\
0 & \text{otherwise},
\end{cases}
\]
and is referred to as an $(n,m,d,k,l,\alpha)$-list UF matrix.

Furthermore, for $0 \leq \alpha, \epsilon \leq 1$, a matrix $M \in \{0,1\}^{m \times n}$ is called an $(n,m,d,k,\epsilon,\alpha)$-\emph{strongly list UF} matrix if it is $(n,m,d,k',\epsilon k',\alpha)$-list UF for every $k' \leq k$.
\end{definition}

\begin{remark}
    An $(n,m,d,k,1,1)$-list UF matrix, corresponding to the  case $l=\alpha=1$, is precisely a $k$-disjunct matrix with constant column weight $d$. More generally, an $(n,m,d,k,l,1)$-list UF matrix, corresponding to the case $\alpha=1$, is equivalent to a (constant column weight) $(k,l)$-list disjunct matrix \cite{ngo2011efficiently,matsumoto2023improved}. The latter generalizes the notion of $k$-disjunct matrices and has been extensively studied in the group testing and coding theory literature.
\end{remark}

We will work with a closely related, but more convenient, notion tailored to our decoding procedure, namely that of $(k,l)$-distinguishable matrices. Throughout the paper, we assume $0 < l < k$ and $n\geq k+l$. 

\begin{definition}\label{def:kld}
A binary matrix $M \in \{0,1\}^{m \times n}$ is called $(k,l)$-\emph{distinguishable} if for any vector $x \in \{0,1\}^n$ with $|\supp(x)| = k$, we have
\[
\left| \left\{ i \in \supp(x) : \exists j \in [m] \text{ s.t. } \supp(M^j) \cap \supp(x) = \{i\} \right\} \right| > k - l.
\]


\end{definition}

Intuitively, this property ensures that at least $k - l + 1$ indices in the support can be uniquely identified (as singletons) from the measurements. This feature is crucial for the first stage of our decoding algorithm; see Section~\ref{sec:usrAlgorithms} for details. 

\begin{example}
Let $n=m=3$, $k=2$, and $l=1$. Consider the matrix
\[
M=
\begin{bmatrix}
1 & 0 & 1 \\
0 & 1 & 1 \\
1 & 1 & 0
\end{bmatrix}.
\]
We claim that $M$ is $(2,1)$-distinguishable.

To verify this, consider any $x\in\{0,1\}^3$ s.t. $|\supp(x)|=2$. There are three possibilities:

\begin{itemize}
    \item If $\supp(x)=\{1,2\}$, then row 1 isolates index $1$ (since $\supp(M^1)\cap\{1,2\}=\{1\}$), and row 2 isolates index $2$.
    \item If $\supp(x)=\{1,3\}$, then row 2 isolates index $3$, and row 3 isolates index $1$.
    \item If $\supp(x)=\{2,3\}$, then row 1 isolates index $3$, and row 3 isolates index $2$.
\end{itemize}


In each case, both elements in $\supp(x)$ are isolated by some row. In other words, every two-column submatrix of $M$ contains a submatrix equal to either
\(\renewcommand{\arraystretch}{0.6}
\begin{bmatrix}
1 & 0 \\[2pt]
0 & 1
\end{bmatrix}
\text{ or } 
\begin{bmatrix}
0 & 1 \\[2pt]
1 & 0
\end{bmatrix},
\)
This idea will be used later in the proof of Proposition~\ref{prop2.1.1}. Hence, $M$ is $(2,1)$-distinguishable.

\end{example}

To handle all sparsity levels up to $k$, we further define the following stronger notion.

\begin{definition}\label{def:strongbasic}
A binary matrix $M \in \{0,1\}^{m \times n}$ is called $(k,\epsilon)$-\emph{strongly distinguishable} if it is $(k', \epsilon k')$-distinguishable for every $k' \leq k$.
\end{definition}

In the following proposition and corollary, we establish the connection between distinguishable matrices and list UF matrices.

\begin{proposition}\label{prop2.1.1}
    A $(n,m,d,k,l, 1)$-list UF matrix is $(k,l)$-distinguishable.
\end{proposition}
\begin{proof}
    Let $M$ be an $(n,m,d,k,l,1)$-list union-free (UF) matrix. Consider any set of $k$ columns of $M$, denoted by $M_{j_1}, \dots, M_{j_k}$. Let $V:=\{j_1,...,j_k\}$, and
\(
M' := M_{V} = [M_{j_1}, \dots, M_{j_k}]
\)
be the corresponding submatrix. 
By Definition~\ref{def:kld}, to show that $M$ is $(k,l)$-distinguishable, it suffices to exhibit a submatrix of $M'$ consisting of at least $k - l + 1$ rows, each equal to a distinct standard basis vector.

We construct such a submatrix iteratively. First, note that any $(n,m,d,k,l,1)$-list UF matrix is also $(n,m,d,k-l,l,1)$-list UF, which  follows from the definition, as any $k-l$ column set can be extended to a $k$ column set by appending $l$ columns.

Let $S \subseteq V$ be any subset of size $l$, for instance
\(
S := \{j_1,...,j_l\}\), and set \( T := V \setminus S.
\)

By the $(n,m,d,k-l,l,1)$-list UF property applied to $(S,T)$, there exist an index $c_1 \in [l]$ and a row index $\lambda_1 \in [m]$ such that
\[
M_{j_{c_1}}[\lambda_1] = 1, \text{ and } M_{j'}[\lambda_1] = 0 \ \text{for all } j'\in \{j_1,...,j_k\} \setminus \{j_{c_1}\}.
\]
Thus, the $\lambda_1$-th row of $M'$ is equal to a standard basis vector, say $e_{i_1}$.

Next, define new subsets
\(
S_2 := \bigl(S \setminus \{j_{c_1}\}\bigr) \cup \{j_{l+1}\},  T_2 := V \setminus S_2.
\)
Applying the same property to $(S_2, T_2)$ yields another row $\lambda_2$ of $M'$ equal to $e_{i_2}$, where $i_2 \neq i_1$ and $\lambda_2\neq \lambda_1$ as equality in either case would result in a collision.

Repeating this procedure, at each step replacing one index in $S$ with a new index from $V \setminus S$, we obtain $k - l + 1$ distinct rows of $M'$, each equal to a distinct standard basis vector.

Therefore, $M'$ contains a submatrix with at least $k - l + 1$ rows forming distinct standard basis vectors, as required.
\end{proof}

Using Proposition~\ref{prop2.1.1}, combined with Definition~\ref{def:basic} and Definition~\ref{def:strongbasic}, we derive: 
\begin{corollary}\label{cor2.1.1}
    A $(n,m,d,k,\epsilon, 1)$-strongly list UF matrix is $(k,\epsilon)$-strongly-distinguishable.
\end{corollary}

In the case of universal $\epsilon$-approximate recovery, we rely on the $(k,\epsilon)$-strongly-distinguishable matrix, while in the case of exact recovery, we rely on a matrix that is $(k',1)$-distinguishable for all $k'\leq k$. Fortunately, a $(k,1)$-distinguishable matrix satisfies this property, which follows directly from Definition~\ref{def:kld} by letting $l=1$. We state this observation as the following proposition, which will be used in our proof:
\begin{proposition}\label{lem:k1d}
    A $(k,1)$-distinguishable matrix is also $(k',1)$-distinguishable for all $k'\leq k$. 
\end{proposition}

The next two corollaries give upper bounds on dimensions of a $(k,\epsilon)$-strongly-distinguishable matrix and a $(k,1)$-distinguishable matrix, which can be proved by 
combining Corollary~\ref{cor2.1.1} and \cite[Corollary~1]{matsumoto2023improved}, as well as Proposition~\ref{prop2.1.1} and \cite[Lemma~2.3]{matsumoto2023improved}: 
\begin{corollary}\label{col:sdm}
    There exists a $(k,\epsilon)$-strongly-distinguishable matrix with $O(k\epsilon^{-1}\log(n/k))$ rows and constant column weight $d=O(\epsilon^{-1}\log(n/k))$. 
\end{corollary}
\begin{corollary}\label{col:dm}
    There exists a $(k,1)$-distinguishable matrix with $O(k^2\log(n/k))$ rows and constant column weight $d=O(k\log(n/k))$. 
\end{corollary}

Having defined $(k,l)$-(strongly) distinguishable matrices, we now fix such a matrix $M$ for use in our constructions. To enable \emph{block decoding} (Section~\ref{sec:bd}), we transform $M$ into the matrix $A'$. However, relying solely on $A'$ leads to the presence of false positives in the recovered support (see Section~\ref{sec:bd} for details). To address this issue, we incorporate an additional matrix $A''$, which is chosen to be an $(n,m,d,k,l,\alpha)$-list union-free matrix for appropriate parameters. This matrix is used in the second stage of the decoding algorithm to filter out false positives.

The overall sensing matrix is constructed as the vertical concatenation
\(
A := [A'; A''].
\)
For simplicity and without loss of generality, we fix $\alpha = \tfrac{1}{2}$ for $A''$, following~\cite{matsumoto2023improved}.

We will make use of known results on the existence and parameters of such matrices. In particular, we restate Corollary~1 and Lemma~2.3 from~\cite{matsumoto2023improved} as the following propositions, which characterize the dimensions of $A''$.

\begin{proposition}\label{lem:suf}
    There exists a $(n,m,d,k,\epsilon,1/2)$-strongly list UF matrix with $O(k\epsilon^{-1}\log(n/k))$ rows and constant column weight $d=O(\epsilon^{-1}\log(n/k))$. 
\end{proposition}
\begin{proposition}\label{lem:uf}
    There exists a $(n,m,d,k,1,1/2)$-list UF matrix with $O(k^2\log(n/k))$ rows and constant column weight $d=O(k\log(n/k))$. 
\end{proposition}

\subsection{Block decoding}\label{sec:bd}
To improve the decoding time, we adopt the idea of \emph{block decoding} from the group testing literature~\cite{cai2017efficient,lee2019saffron,guruswami2023noise}. This approach partitions the sensing matrix $A$ and the measurement vector $y$ into vertical blocks. Specifically, for a parameter $L$, we write
\[
A := [A_1; A_2; \dots; A_L], \quad
y := [y_1; y_2; \dots; y_L],
\]
where each $A_i$ corresponds to a block of rows and $y_i$ is the associated subvector of measurement results. The decoder processes each block $y_i$ independently, and aggregates the information across all blocks to recover the support.

In our universal support recovery schemes, we build on an algorithm  proposed in~\cite{lee2019saffron}, called \textsc{SAFFRON},  which enables fast decoding in nonadaptive group testing. A key idea in \textsc{SAFFRON} is to identify \emph{singletons}, i.e., measurements that correspond to exactly one defective item, and then directly recover their indices. Such singletons can be decoded in near-constant time. 

To facilitate this, \textsc{SAFFRON} employs a \emph{signature matrix} $U$ (see Section~3.1 of~\cite{lee2019saffron}):
\[
U := [U^{(1)}; \overline{U^{(1)}}] =
\begin{bmatrix}
    b_1 & b_2 & \cdots & b_n \\
    \overline{b_1} & \overline{b_2} & \cdots & \overline{b_n}
\end{bmatrix},
\]
where $b_i \in \{0,1\}^{\lceil \log_2 n \rceil}$ is the binary representation of $i-1$, and $\overline{b_i}$ denotes its bitwise complement. This matrix has $2\lceil \log_2 n \rceil$ rows and $n$ columns, with each column having weight exactly $\lceil \log_2 n \rceil$.

A crucial property of $U$ is that the union (bitwise OR) of any two or more distinct columns has weight strictly greater than $\lceil \log_2 n \rceil$, whereas each individual column has weight exactly $\lceil \log_2 n \rceil$. This allows us to distinguish singleton measurements from those involving multiple coordinates (in the group testing setting).

In our 1bCS setting, we define a \emph{singleton} to be an index $s \in \supp(x)$ such that there exists a row $M^i$ of the measurement matrix satisfying
\[
\supp(M^i) \cap \supp(x) = \{s\}.
\]
Identifying such singletons is the key step in our decoding algorithm.

To enable efficient singleton detection, we apply a \emph{$U$-magnification} process to the matrix $M$: each nonzero entry of $M$ is replaced by the corresponding column of $U$, while each zero entry is replaced by the zero vector $0^{2\lceil \log_2 n \rceil}$. This transformation produces a new sensing matrix that supports fast block decoding, as formalized below.

\begin{definition}\label{def:umm}
    Given a binary matrix $M \in \{0,1\}^{m' \times n}$ and a signature matrix $U \in \{0,1\}^{u \times n}$, we define the \emph{$U$-magnified matrix} of $M$, denoted by $A' \in \{0,1\}^{um' \times n}$, as follows.

For each row index $i \in [m']$, replace the $i$-th row $M^i$ of $M$ with a block of $u$ rows. Specifically, for each column index $j \in [n]$, the corresponding block entry is given by
\[
A'[(i-1)u + 1 : iu,\, j] =
\begin{cases}
U_j & \text{if } M_{ij} = 1,\\
0^u & \text{if } M_{ij} = 0,
\end{cases}
\]
where $U_j$ denotes the $j$-th column of $U$.
\end{definition}

\begin{example}\label{exp:matrices}
Let

\[
M=\begin{bmatrix}
0&0&0&1&0&1&0&0\\
1&0&0&0&1&0&0&1
\end{bmatrix}, \]
\[
U=\begin{bmatrix}
0&0&0&0&1&1&1&1\\
0&0&1&1&0&0&1&1\\
0&1&0&1&0&1&0&1\\
1&1&1&1&0&0&0&0\\
1&1&0&0&1&1&0&0\\
1&0&1&0&1&0&1&0
\end{bmatrix}
,
\]

Then the $U$-magnified matrix of $M$, denoted by $A'$, is
\[
A'=\begin{bmatrix}
0&0&0&0&0&1&0&0\\
0&0&0&1&0&0&0&0\\
0&0&0&1&0&1&0&0\\
0&0&0&1&0&0&0&0\\
0&0&0&0&0&1&0&0\\
0&0&0&0&0&0&0&0\\
0&0&0&0&1&0&0&1\\
0&0&0&0&0&0&0&1\\
0&0&0&0&0&0&0&1\\
1&0&0&0&0&0&0&0\\
1&0&0&0&1&0&0&0\\
1&0&0&0&1&0&0&0
\end{bmatrix}
.
\]
\end{example}

The $U$-magnification process enables efficient identification of singletons via block decoding. Specifically, in the first stage of the decoding algorithm, we partition the result vector $y':=\nz(A'x)$ into blocks of length $2\lceil \log_2 n\rceil$. Each block is examined to determine whether it matches a column of the signature matrix $U$. If a block matches the $j$-th column of $U$, then the index $j$ is added to the set $S$. The elements of $S$ are referred to as \emph{potential singletons}.

We emphasize the term ``potential'' since this procedure may introduce \emph{false positives}, i.e., indices that do not belong to the true support but are mistakenly identified due to collisions in the measurements. The following example illustrates this phenomenon.

\begin{example}\label{exp:FP}
Let $n=8$ and $x=[1;0;-1;0;0;0;1;0]$, so that $\supp(x)=\{1,3,7\}$. Let the signature matrix $U$ be as defined in Example~\ref{exp:matrices}. Consider a measurement row $M^1=[1,0,1,0,0,0,1,1]$. After $U$-magnification, we obtain
\[
A^{M^1}=
\begin{bmatrix}
0 & 0 & 0 & 0 & 0 & 0 & 1 & 1 \\
0 & 0 & 1 & 0 & 0 & 0 & 1 & 1 \\
0 & 0 & 0 & 0 & 0 & 0 & 0 & 1 \\
1 & 0 & 1 & 0 & 0 & 0 & 0 & 0 \\
1 & 0 & 0 & 0 & 0 & 0 & 0 & 0 \\
1 & 0 & 1 & 0 & 0 & 0 & 1 & 0 
\end{bmatrix}.
\]
Then $A^{M^1}x=[1;0;0;0;1;1]=\nz(A^{M^1}x)$, which matches the $5$-th column of $U$. Consequently, the decoder incorrectly adds the coordinate $5$ to $S$, even though $5\notin \supp(x)$. 
\end{example}

\begin{remark}
    We mentioned above the issue of false positives; however, false negatives, i.e. indices that belong to the true support but are not identified by the decoding algorithm, do not require separate consideration. By the properties of the distinguishable matrices, the set $S$ will include all elements in $\supp(x)$ in the exact-recovery setting and more than $1-\epsilon$ portion of $\supp(x)$ in the $\epsilon$-approximate recovery setting. Details will be shown in the proofs of the main theorems. 
\end{remark}

To eliminate such false positives, the second stage of the decoding algorithm refines $S$ as follows. For each $s\in S$, we compare the $s$-th column of $A''$ with $y'':=\nz(A''x)$. If the intersection is sufficiently large, $s$ is retained; otherwise, it is removed. This filtering step ensures that only true support elements remain.

\subsection{EDOCS schemes: universal support recovery}\label{sec:usrAlgorithms}
In this section, we formally present our sensing matrices for the EDOCS-AA (Algorithm~\ref{alg:edocsaa}) and EDOCS-AE (Algorithm~\ref{alg:edocsae}), and their corresponding decoding algorithm (Algorithm~\ref{alg:unidec}). Also, we present the main theorems (Theorem~\ref{thm:aa}, Theorem~\ref{thm:ae}), for universal support recovery. Note that in the construction algorithms, we replaced the parameters $\epsilon$ with $\frac{\epsilon}{2}$, the reason will be shown in the proof of Theorem~\ref{thm:aa}.
 
\begin{algorithm}[!]
\caption{EDOCS-AA: Sensing matrix}\label{alg:edocsaa}
\begin{algorithmic}[1]
    \Require Signal vector dimension $n$, upper bound on the support of signal vector $k$, the signature matrix $U$, and parameter $\epsilon$.  
    \State By Corollary~\ref{col:sdm}, take a $(k,\epsilon/2)$-strongly-distinguishable matrix $M$ with $m'=O(k\epsilon^{-1}\log(n/k))$ rows and constant column weight $d'=O(\epsilon^{-1}\log(n/k))$. 
    \State Let $A'$ be the $U$-magnified matrix (Definition~\ref{def:umm}) of $M$. 
    \State By Proposition~\ref{lem:suf}, take a $(n,m'',d'',k,\epsilon/2,1/2)$-strongly list UF matrix $A''$ with $m''=O(k\epsilon^{-1}\log(n/k))$ rows and constant column weight $d''=O(\epsilon^{-1}\log(n/k))$. 
    \State Let $A:=[A';A'']$, the vertical concatenation of $A'$ and $A''$. 
     
    \State \Return $A$
\end{algorithmic}
\end{algorithm}

\begin{algorithm}[!]
\caption{EDOCS-AE: Sensing matrix}\label{alg:edocsae}
\begin{algorithmic}[1]
    \Require Signal vector dimension $n$, upper bound on the support of signal vector $k$, and the signature matrix $U$.  
    \State By Corollary~\ref{col:dm}, take a $(k,1)$-distinguishable matrix $M$ with $m'=O(k^2\log(n/k))$ rows and constant column weight $d'=O(k\log(n/k))$. 
    \State Let $A'$ be the $U$-magnified matrix (Definition~\ref{def:umm}) of $M$. 
    \State By Proposition~\ref{lem:uf}, take a $(n,m'',d'',k,1,1/2)$-list UF matrix $A''$ with $m''=O(k^2\log(n/k))$ rows and constant column weight $d''=O(k\log(n/k))$. 
    \State Let $A:=[A';A'']$, the vertical concatenation of $A'$ and $A''$. 
    \State \Return $A$
\end{algorithmic}
\end{algorithm}

\begin{algorithm}[!]
\caption{EDOCS-AA and EDOCS-AE: Decoding algorithm \label{alg:unidec}}
\begin{algorithmic}[1]
    \Require Notations $n$, $k$, $U$, $A=[A';A'']$, $d''$ from Algorithm~\ref{alg:edocsaa} (for EDOCS-AA) or Algorithm~\ref{alg:edocsae} (for EDOCS-AE). The result vector $y=[y';y'']$, where $y'=\nz(A'x)$ and $y''=\nz(A''x)$.
    \State Initialize empty set $S$. 
    \State Partition $y'=[y^{(1)};y^{(2)};...;y^{(m')}]$, where each $y^{(i)}$ is of length $2\lceil \log_2 n\rceil$. 
    \For{$i=1,2,...,m'$} \Comment{In the \textbf{first stage}, we find all suspected indices}
    \If {$|\supp(y^{(i)})|=\lceil\log_2 n \rceil$} 
    \State Let $s_i$ be the integer corresponding to the binary representation of the first $\lceil \log_2 n\rceil$ entries of $y^{(i)}$. \State Include $s_i+1$ into the set $S$. 
    \EndIf
    \EndFor
    \For{$i$ in $S$} \Comment{In the \textbf{second stage}, we iterate over all suspected indices and eliminate if necessary}
    \If{$|\supp(A''_i)\cap\supp(y'')|< d''/2$}
    \State $S\leftarrow S-\{i\}$. 
    \EndIf
    \EndFor
    \State \Return $S$
\end{algorithmic}
\end{algorithm}

For universal $\epsilon$-approximate support recovery 1bCS, we have the following result: 

\begin{theorem}~\label{thm:aa}
    \emph{(EDOCS-AA)} Given $\epsilon>0$, there exists a one-bit compressed sensing matrix $A\in\mathbb{R}^{m\times n}$ for universal $\epsilon$-approximate support recovery of all $k$-sparse signal vectors with $m=O(k\epsilon^{-1}\log(n/k)\log n)$ measurements (Algorithm~\ref{alg:edocsaa}), and a recovery algorithm (Algorithm~\ref{alg:unidec}) with  runtime $D=O(\epsilon^{-1}m)$.
\end{theorem}

\begin{proof}
We use the notation from Algorithm~\ref{alg:edocsaa} and Algorithm~\ref{alg:unidec}, unless otherwise specified.

\emph{First, we prove the correctness of the scheme, i.e. The $\epsilon$-approximate support recovery criterion \eqref{eqn:asr} holds.}

In Algorithm~\ref{alg:unidec}, $A = [A'; A'']$ is the sensing matrix and $y = [y'; y'']$ is the result vector. The vector $y'$ is partitioned as $y' = [y^{(1)}; y^{(2)}; \dots; y^{(m')}]$, and $U$ denotes the signature matrix. In Algorithm~\ref{alg:edocsaa}, $M$ is a $(k,\frac{\epsilon}{2})$-strongly distinguishable matrix, and $A'$ is obtained by $U$-magnification of $M$.

By construction,
\(
y' = \nz(A'x) = [y^{(1)}; y^{(2)}; \dots; y^{(m')}],
\)
where for each $i \in [m']$,
\[
y^{(i)} = \nz\left(\sum_{j \in \supp(x)\cap \supp(M^i)} x_jU_j\right).
\]

Let $S'$ denote the output of the first stage of the decoding algorithm, which is to be distinguished from the final output $S$. We show that $S'$ captures a large fraction of $\supp(x)$, namely,
\begin{equation}\label{eqn1}
    |S' \cap \supp(x)| > \left(1 - \frac{\epsilon}{2}\right)|\supp(x)|.
\end{equation}

Since $|\supp(x)| \leq k$ and $M$ is $(k,\frac{\epsilon}{2})$-strongly distinguishable, more than a $(1 - \frac{\epsilon}{2})$ fraction of the indices in $\supp(x)$ appear as singletons. That is,


\begin{equation}\label{eqn2}
\begin{aligned}
&\Bigl| \Bigl\{ j \in \supp(x) :\; 
\exists i \in [m'] \text{ s.t. }
\supp(x) \cap \supp(M^i) = \{j\} \Bigr\} \Bigr| > \left(1 - \frac{\epsilon}{2}\right) |\supp(x)|.
\end{aligned}
\end{equation}

Hence, for each $j$ in the set on the left hand side of~\eqref{eqn2}, there exists an $i$ such that $y^{(i)} = \nz(x_jU_j)$. By Algorithm~\ref{alg:unidec} and the properties of the signature matrix $U$, the index $j$ is correctly decoded and added to $S'$, and thus~\eqref{eqn1} follows.

Going into the second stage of the decoding algorithm, denote
\[
    S_0:=\{i\in[n]:|\supp(A_i'')\cap\supp(y'')|\geq d''/2\},
\]
\[
    S_0':=\{i\in[n]:|\supp(A_i'')\cap\bigcup_{j\in\supp(x)}\supp(A''_j)|\geq d''/2\}.
\]

Since $y''=\nz\left(\sum_{j\in\supp(x)}A_j''\right)$, it follows that
\[\supp(y'')\subseteq \bigcup_{j\in\supp(x)}\supp(A''_j)\Rightarrow S_0\subseteq S_0'. \] 
We then show that
\begin{equation}\label{eqn3}
\begin{aligned}
    & |S_0 \setminus \supp(x)| \leq \frac{\epsilon}{2}|\supp(x)|, \\
    & |\supp(x) \setminus S_0| \leq \frac{\epsilon}{2}|\supp(x)|.
\end{aligned}
\end{equation}

Consider the set $[n]\setminus \supp(x)$. Let $S_a$ be its subset such that $|S_a|=\frac{\epsilon}{2}|\supp(x)|$ (for simplicity, suppose this is an integer). Applying the $(n,m'',d'',k,\epsilon/2,1/2)$-strongly list union-free property of $A''$ on $S_a$ and $\supp(x)$, $S_a$ contains at least one element not in $S_0'$, and thus not in $S_0$. Since $S_a$ can be chosen arbitrarily, $[n]\setminus \supp(x)$ contains at most $\frac{\epsilon}{2}|\supp(x)|$ elements in $S_0$, i.e. $|S_0 \setminus \supp(x)| \leq \frac{\epsilon}{2}|\supp(x)|$. 

For the second part of \eqref{eqn3}, let $S_c$ be a subset of $\supp(x)$ of size $\frac{\frac{\epsilon}{2}}{1+\frac{\epsilon}{2}}|\supp(x)|$ and let $T_c:=\supp(x)\setminus S_c$. Apply the $(n,m'',d'',k,\epsilon/2,1/2)$-strongly list union-free property of $A''$ on $S_c$ and $T_c$, there exists at least one element $s\in S_c$ such that \[\left|\supp(A''_s)\cap\bigcup_{j\in\supp(x)\setminus{\{s\}}}\supp(A''_j)\right|< d''/2. \]

This means, the $s$-th column of $A''$ contains $\geq d''/2$ indices at which the entries are nonzero, while the $t$-th column of $A''$ for all $t\in\supp(x)\setminus\{s\}$ at those indices have zero entries. Since $y''=\nz\left(\sum_{j\in\supp(x)}A_j''\right)$, it follows that \[|\supp(A_s'')\cap\supp(y'')|\geq d''/2, \]
i.e. $s\in S_0$. Since $S_c$ can be chosen arbitrarily, by repeating the process above, each time selecting a different $S_c\subseteq \supp(x)$ that does not contain the previously chosen $s\in\supp(x)$ (Also observe that
$1-\frac{\frac{\epsilon}{2}}{1+\frac{\epsilon}{2}}
>1-\frac{\epsilon}{2}$), 
we can identify more than $\left(1-\frac{\epsilon}{2}\right)|\supp(x)|$ elements of $\supp(x)$ that also belong to $S_0$.
Hence $|\supp(x) \setminus S_0| \leq \frac{\epsilon}{2}|\supp(x)|$, and \eqref{eqn3} is proved. 

Continuing, note that in Algorithm~\ref{alg:unidec}, instead of iterating over all $n$ columns of $A''$, we only iterate over columns indexed by $S'$, so $S = S_0 \cap S'$. Combining~\eqref{eqn1} and~\eqref{eqn3}, we get

\begin{equation*}
\begin{aligned}
    & |S \setminus \supp(x)| \leq \epsilon|\supp(x)|, \\
    & |\supp(x) \setminus S| \leq \epsilon|\supp(x)|.
\end{aligned}
\end{equation*}
as desired.

\emph{Next, we analyze the number of measurements and the decoding complexity.}

The total number of measurements is $m = |y'| + |y''|$. The term $|y'|$ equals the number of rows of a $(k,\epsilon)$-strongly distinguishable matrix multiplied by $2\lceil \log_2 n \rceil$, which is
\[
|y'| = O\bigl(k\epsilon^{-1}\log(n/k)\log n\bigr).
\]

The term $|y''|$ corresponds to the number of rows of a $(n,m'',d'',k,\epsilon/2,1/2)$-strongly list union-free matrix, which is
\(
|y''| = O\bigl(k\epsilon^{-1}\log(n/k)\bigr).
\)

Thus, $|y'|$ dominates $|y''|$, and
\[
m = O\bigl(k\epsilon^{-1}\log(n/k)\log n\bigr).
\]

The decoding complexity is the sum of the costs of the two stages. The first stage processes all entries of $y'$, taking time $O(|y'|)=O(m)$. The second stage iterates over columns indexed by $S'$, with complexity
\[
|S'| \cdot d'' \leq m' \cdot d'' 
= O\bigl(k\epsilon^{-1}\log(n/k)\bigr)\cdot O\bigl(\epsilon^{-1}\log(n/k)\bigr)= O(\epsilon^{-1} m).\]


Therefore, the overall decoding complexity is
\[
D = O(\epsilon^{-1} m).
\]
\end{proof}

Similarly, for the universal exact support recovery 1bCS, we have the following result:  
\begin{theorem}~\label{thm:ae}
    \emph{(EDOCS-AE)} There exists a one-bit compressed sensing matrix $A\in\mathbb{R}^{m\times n}$ for universal exact support recovery of all $k$-sparse signal vectors with $m=O(k^2\log(n/k)\log n)$ measurements (Algorithm~\ref{alg:edocsae}), and a recovery algorithm (Algorithm~\ref{alg:unidec}) with runtime $D=O(km)$. 
\end{theorem}

\begin{proof}
The proof essentially follows from that of Theorem~\ref{thm:aa}. 
Define $S'$ as in the proof of Theorem~\ref{thm:aa}, and note that the matrix $M$ in Algorithm~\ref{alg:edocsae} is $(k',1)$-distinguishable for every $k'\leq k$, the first-stage argument in the proof of Theorem~\ref{thm:aa} gives
\(\supp(x)\subseteq S'.\)

Define $S_0$ as in the proof of Theorem~\ref{thm:aa}. Since an $(n,m'',d'',k,1,1/2)$-list UF matrix is also an $(n,m'',d'',k',1,1/2)$-list UF matrix for every $k'\leq k$ (directly follows from the definition), the second-stage argument in the proof of Theorem~\ref{thm:aa} gives
\(
S_0=\supp(x).
\)

Since the final output is $S=S_0\cap S'$, it follows that
\[
S=\supp(x),
\]
which proves the correctness of the scheme.

The measurement and decoding complexity bounds follow from the same calculations as in the proof of Theorem~\ref{thm:aa}, with the corresponding parameters of the present construction.
\end{proof}





\section{EDOCS: Probabilistic exact support recovery}\label{sec:RSR}
In this section, we present the EDOCS scheme for probabilistic exact support recovery, which we call EDOCS-EE, where ``EE'' stands for ``for Each, Exact.''  

Our approach leverages the fast binary splitting algorithm from the group testing literature~\cite{price2020fast,cheraghchi2020combinatorial,wang2023quickly,price2023fast,li2025noisy}. We will briefly describe this algorithm, explain how it connects to the 1bCS problem, and show that it can be adapted to our setting in a highly efficient manner. 

Without loss of generality, we assume throughout this section that $n$ is a power of $2$. For general $n$, we can append dummy coordinates to obtain a dimension equal to the nearest power of $2$. This modification does not affect the asymptotic scaling of the measurement complexity or the decoding complexity, since the modified dimension differs from $n$ by at most a constant factor and the complexities depend on $n$ only through logarithmic terms.

\subsection{Fast binary splitting algorithm: a summary}\label{sec:fbs}

Following~\cite[Section~III]{wang2023quickly}, we assume throughout this subsection, without loss of generality, that $k$ is a power of $2$. The parameters $n$ and $k$ are inherited from the original group testing formulation and will be used analogously in our 1bCS framework.

The fast binary splitting algorithm for nonadaptive group testing constructs the test matrix based on a complete binary tree of depth $\log_2 n$, with levels indexed from $0$ to $\log_2 n$. Each node in the tree represents a subset of items, and each item is associated with a binary string of length $\log_2 n$, ranging from $[00\ldots0]$ to $[11\ldots1]$. We refer to this binary tree as the \emph{associated binary tree} (ABT) throughout the paper.

The root node (level $0$) corresponds to the entire set of items. Its two children at level $1$, labeled $0$ and $1$, represent the subsets of items whose binary representations begin with $0$ and $1$, respectively. Recursively, each node corresponds to a prefix, and its two children are obtained by appending one additional bit to this prefix. For example, the nodes $00$ and $01$ correspond to subsets of items whose binary representations begin with $00$ and $01$, while the nodes $10$ and $11$ correspond to subsets beginning with $10$ and $11$. This recursive partitioning continues until level $\log_2 n$, where each leaf node corresponds to a single item.

We adopt the construction from~\cite{wang2023quickly}, which consists of two phases:

\paragraph{Grow-and-prune phase}
This phase uses $16k\log_2(n/k)$ tests. For each level from $\log_2(k)+1$ to $\log_2 n$, $16k$ tests are performed, and each node at that level (all items represented by this node) is assigned independently and uniformly at random to one of these tests. This phase aims to identify all candidate items, possibly including false positives.  

\paragraph{Leaf-trimming phase}
This phase uses $16k\log_2 k $ tests, partitioned into $\log_2 k $ groups of $16k$ tests each. For every group, each node at the final level is assigned independently and uniformly at random to one of the $16k$ tests. Conceptually, this phase can be viewed as extending the tree by $\log_2 k$ additional levels, where each node has only one child. Its purpose is to eliminate false positives generated in the first phase.  

We omit the details of the decoding procedure and the error analysis and summarize the performance guarantee in the following lemma:

\begin{lemma}\label{lem:error1}
    (Theorem 4 in~\cite{wang2023quickly}) For the nonadaptive group testing problem, the fast binary splitting algorithm in~\cite{wang2023quickly} will return an erroneous set (i.e. a set containing false positives and/or false negatives) with probability $<e^{-k}+n^{-k}+5k^{-3}$, with decoding complexity asymptotically the same as the number of tests, i.e. $O(k\log n)$.  
\end{lemma} 

\subsection{Connection of Group Testing to One-Bit Compressed Sensing}\label{sec:gt1bcs}

In group testing, the goal is to recover a binary vector $x$ from measurements of the form $y = A \circ x$, where $A$ is a binary matrix and the operator ``$\circ$'' replaces the standard addition and multiplication with logical OR and AND, respectively. In contrast, in our one-bit compressed sensing (1bCS) setting, our objective is to recover the support of a real vector $x \in \mathbb{R}^n$ from measurements of the form $y = \nz(Ax)$.

To relate the two models, given a sensing matrix $A$ and signal $x$ in the 1bCS setting, define $A' := \nz(A)$ and $x' := \nz(x)$. Let $y' := A' \circ x'$ denote the corresponding group testing outcome. Then, for any $i \in [m]$, if $y'_i = 0$, it necessarily follows that $y_i = 0$. On the other hand, if $y'_i = 1$, then typically $y_i = 1$, except in the case where $A^i \cdot x = 0$. We refer to this event, where $(A')^i \circ x' = 1$ but $A^i \cdot x = 0$, as an \emph{accidental zero}.

Therefore, the group testing outcome $y'$ can be used as a proxy for the 1bCS measurement $y$, provided that accidental zeros can be avoided. A standard approach, suggested in~\cite{matsumoto2023improved}, is to repeat each row of $A$ by $k$ times and modify the nonzero entries appropriately (e.g., via totally invertible matrices; see Definition~\ref{def:tim}). However, this introduces an undesirable multiplicative factor of $k$ in the number of measurements.

In the for-each setting, this overhead can be significantly reduced. By designing the sensing matrix $A$ based on the fast binary splitting scheme, we ensure that, with high probability, each measurement $A^i$ intersects $\supp(x)$ in only $O(\log k)$ coordinates. Consequently, repeating each row only $O(\log k)$ times suffices to avoid accidental zeros. Further details of this construction and analysis are provided in Section~\ref{sec:bib}.

\paragraph{Remark.}
Among various group testing schemes, we adopt the fast binary splitting algorithm due to its efficiency. To the best of our knowledge, it achieves both optimal measurement complexity and decoding time of $O(k \log n)$ in the noiseless setting. Moreover, its randomized construction naturally aligns with the analysis of the multivariate hypergeometric distribution (see Section~\ref{sec:bib}), which plays a key role in our performance guarantees.

\subsection{Bounding the intersection between the support of signal vector and that of any measurement}\label{sec:bib}

Recall from Section~\ref{sec:fbs} that the fast binary splitting algorithm consists of $\log_2 n$ steps. At each step, the algorithm operates on a specific level of the ABT, where every node at that level is assigned independently and uniformly at random to one of the $16k$ tests.

Throughout this section, we assume without loss of generality that
$|\supp(x)|=k$. Under the probabilistic model introduced in
Section~\ref{sec:1bCS}, the support of the signal vector is chosen uniformly
at random from all ${n\choose k}$ subsets of $[n]$. Equivalently, in the
corresponding group testing formulation, the $k$ defective items are chosen
uniformly at random from all subsets of $k$ items among the $n$ items.

As discussed in Section~\ref{sec:gt1bcs}, in order for the proposed 1bCS
scheme to achieve both correctness and computational efficiency, under the group testing setting, it is
necessary to guarantee that the maximum number of defective items
participating in any test of the underlying fast binary splitting algorithm is
$O(\log k)$ with high probability.

To provide additional intuition and facilitate our analysis, we associate each node of the ABT with the number of defective items contained in its corresponding subtree. At the leaf level of the ABT, each node represents a single item. Therefore, exactly $k$ leaf nodes are assigned the value $1$, corresponding to the defective items, while the remaining leaf nodes are assigned the value $0$. The value associated with each internal node is defined as the total number of defective leaves in its subtree.

Consider a fixed level of the ABT containing $u$ nodes. Since the fast binary splitting algorithm begins at level $\log_2(k)+1$ of the ABT, we always have $u\geq k$. Each node corresponds to a subtree, and we define its \emph{capacity} as the number of leaves contained in that subtree. Since $n$ is
assumed to be a power of $2$, the ABT is perfect, and every node at
the same level has identical capacity
\(
c=\frac{n}{u}.
\)

Let $(Y_1,\ldots,Y_u)$ denote the numbers of defective items contained in the
$u$ nodes at this level, we have
\begin{equation}\label{eqn:multi}
(Y_1,\ldots,Y_u)
\sim
\operatorname{MultiHypergeometric}
\left(
n,k;\underbrace{c,c,\ldots,c}_{u\ \text{times}}
\right).
\end{equation}

\begin{example}
    Let $n=16$ and $k=6$. We index the items from $1$ to $16$ and assume that the defective items are located at $\{5,6,8,12,15,16\}$. The following figure illustrates the ABT, where each node is annotated with the number of defective items contained in its subtree.

\begin{figure}[H]
    \centering
    \includegraphics[height=8cm, width=15cm]{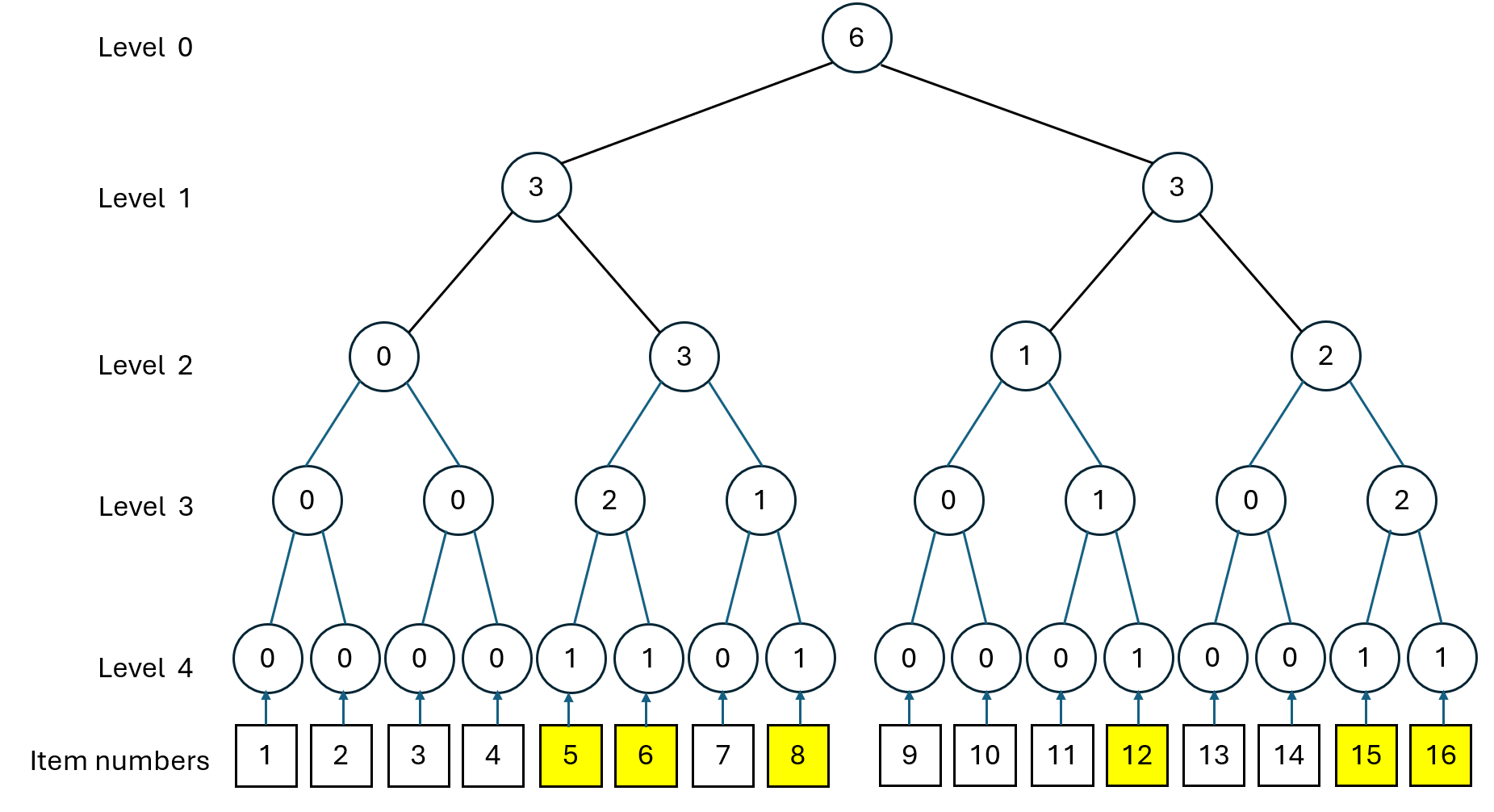}
    \caption{An example of binary tree with occupancy numbers}
\end{figure}


At level $2$ of the ABT, each node has capacity $4$. Suppose one test is assigned nodes $2$ and $3$ (with node indices ordered from left to right), while another test is assigned nodes $1$ and $4$. Then, the corresponding tests contain $4(=3+1)$ and $2(=0+2)$ defective items, respectively.

At this level, we have $u=4$, and the vector $(Y_1,Y_2,Y_3,Y_4)=(0,3,1,2)$ can be viewed as a realization of the multivariate hypergeometric distribution
\[
\operatorname{MultiHypergeometric}(16,6;4,4,4,4).
\]

\end{example}

The following lemma formalizes the above discussion and establishes the desired bound.

\begin{lemma}\label{lem:error2}
Let
\(
k_0:=10\log_2 k .
\)
Under the fast binary splitting construction of~\cite{wang2023quickly}, the
probability that there exists a test containing more than $k_0$ defective
items is at most
\(
20\log_2(n)k^{-9}.
\)
\end{lemma}

\begin{proof}
We adopt the notation and terminology introduced in the preceding discussion.

Fix one level of the binary tree associated with the fast binary splitting construction containing $u\geq k$ nodes. Let $X_i$ denote
the number of defective items contained in test $i$ under the original
construction. Since every node is assigned independently and uniformly to one
of the $16k$ tests,
\[
X_i=\sum_{j=1}^{u}Y_jI_j,
\]
where $\{I_j\}_{j=1}^{u}$ are independent Bernoulli random variables with
parameter $(16k)^{-1}$.

It remains to bound the tail probability of $X_i$. Define the probability generating function
\[
G(z):=\mathbb{E}[z^{X_i}],\quad z>1 .
\]

Conditioning on $(Y_1,\ldots,Y_u)$ gives
\[
\begin{aligned}
G(z)
&=
\mathbb{E}
\left[
z^{\sum_{j=1}^{u}Y_jI_j}
\right]
=
\mathbb{E}
\left[
\prod_{j=1}^{u}
\left(
1-\frac1{16k}
+\frac{z^{Y_j}}{16k}
\right)
\right].
\end{aligned}
\]

Expanding the product,
\[
G(z)
=
\sum_{A\subseteq[u]}
\left(\frac1{16k}\right)^{|A|}
\left(1-\frac1{16k}\right)^{u-|A|}
\mathbb{E}
\left[
z^{\sum_{j\in A}Y_j}
\right].
\]





For any fixed subset $A\subseteq [u]$, using~\eqref{eqn:multi}, we have
\[
\sum_{j\in A}Y_j
\sim
\operatorname{Hypergeometric}
\left(
n,c|A|,k
\right).
\]

Let
\[
(B_1,\ldots,B_u)\sim
\operatorname{Multinomial}
\left(k;\frac1u,\ldots,\frac1u\right),
\]

so that
\[
\sum_{j\in A}B_j
\sim
\operatorname{Bin}
\left(
k,\frac{|A|}{u}
\right).
\]

We next compare the two random variables. Consider a population of $n=cu$
positions, where the positions corresponding to the boxes in $A$ form a
subset of size $c|A|$. The random variable $\sum_{j\in A}Y_j$ counts the
number of sampled positions from this subset when $k$ positions are chosen
uniformly without replacement. On the other hand,
$\sum_{j\in A}B_j$ counts the number of successes in $k$ independent trials,
where each trial is successful with probability
\[
\frac{c|A|}{n}=\frac{c|A|}{cu}=\frac{|A|}{u}.
\]

By the standard comparison theorem between sampling without replacement and
sampling with replacement~\cite{hoeffding1963probability}, 
for any nondecreasing convex function $\phi$,
\[
\mathbb{E}\left[\phi\left(\sum_{j\in A}Y_j\right)\right]
\leq
\mathbb{E}\left[\phi\left(\sum_{j\in A}B_j\right)\right].
\]

Since $z^x$ is nondecreasing and convex for $z\geq 1$, taking
$\phi(x)=z^x$ gives
\[
\mathbb{E}
\left[
z^{\sum_{j\in A}Y_j}
\right]
\leq
\mathbb{E}
\left[
z^{\sum_{j\in A}B_j}
\right].
\]

Finally, since
\[
\mathbb{E}
\left[
z^{\sum_{j\in A}B_j}
\right]
=
\left(
1+\frac{|A|}{u}(z-1)
\right)^k, 
\]

we obtain
\[
\mathbb{E}
\left[
z^{\sum_{j\in A}Y_j}
\right]
\leq
\left(
1+\frac{|A|}{u}(z-1)
\right)^k .
\]

Therefore, $G(z)$ is upper bounded by
\[
\sum_{j=0}^{u}
{u\choose j}
\left(\frac1{16k}\right)^j
\left(1-\frac1{16k}\right)^{u-j}
\left(
1+\frac{j}{u}(z-1)
\right)^k .
\]

Equivalently,
\[
G(z)
\leq
\mathbb{E}_{M}
\left[
\left(
1+\frac{M}{u}(z-1)
\right)^k
\right],
\]

where
\[
M\sim
\operatorname{Bin}
\left(u,\frac1{16k}\right).
\]

Using $1+x\leq e^x$,
\[
G(z)
\leq
\mathbb{E}_{M}
\left[
\exp
\left(
\frac{kM}{u}(z-1)
\right)
\right].
\]

The moment generating function of $M$ gives
\[
G(z)
\leq
\exp
\left(
\frac{u}{16k}
\left(
\exp\left(\frac{k(z-1)}{u}\right)-1
\right)
\right).
\]

Let $\alpha=\frac{u}{k}\geq1$, then
\[
G(z)
\leq
\exp
\left(
\frac{\alpha}{16}
\left(
e^{(z-1)/\alpha}-1
\right)
\right).
\]

Choosing $z=2$,
\[
G(2)
\leq
\exp
\left(
\frac{\alpha}{16}(e^{1/\alpha}-1)
\right).
\]

Since $\alpha(e^{1/\alpha}-1)$ is decreasing for $\alpha\geq1$, we have 
\[
\alpha(e^{1/\alpha}-1)\leq e-1,
\]

and therefore

\[
G(2)
\leq
\exp\left(\frac{e-1}{16}\right)
<1.2 .
\]

By Markov's inequality,
\[
\Pr(X_i\geq r)
\leq
\frac{G(2)}{2^r}.
\]

Taking $r=10\log_2k$, we obtain
\[
\Pr(X_i\geq10\log_2k)
\leq
1.2k^{-10}.
\]

Applying a union bound over all $16k$ tests,
\[
\Pr
\left(
\max_{i\in[16k]}X_i\geq10\log_2k
\right)
\leq
16k\cdot1.2k^{-10}
\leq
20k^{-9}.
\]

Finally, the above argument applies to every level of the binary tree. Since
there are $\log_2(n)$ levels, another union bound finishes the proof. 
\end{proof}

\subsection{EDOCS scheme: probabilistic exact support recovery}\label{sec:EEE}
Before presenting the EDOCS-EE scheme, we introduce several definitions related to totally invertible matrices that are central to our construction.

\begin{definition}\label{def:tim}
A matrix $M\in\mathbb{R}^{m\times n}$ is said to be \emph{totally invertible} if every square submatrix of $M$ is invertible.
\end{definition}

\begin{definition}
Given a binary vector $v\in \{0,1\}^n$, its \emph{associated sub-totally invertible matrix with $l$ rows} (AsTIM-$l$) is a matrix $V\in\mathbb{R}^{l\times n}$ such that $V_{ij}\neq 0$ if and only if $v_j=1$, and the submatrix of $V$ restricted to the nonzero columns is totally invertible.
\end{definition}

\begin{example}\label{exp:astim}
For the vector $v=[0;1;1;0;1;0;0;1]$, one explicit construction of an AsTIM-$3$ matrix, based on the Hilbert matrix, is given by
\[
V={
   \renewcommand{\arraystretch}{1.2} 
\begin{bmatrix}
0 & 1 & \frac{1}{2} & 0 & \frac{1}{3} & 0 & 0 & \frac{1}{4} \\
0 & \frac{1}{2} & \frac{1}{3} & 0 & \frac{1}{4} & 0 & 0 & \frac{1}{5} \\
0 & \frac{1}{3} & \frac{1}{4} & 0 & \frac{1}{5} & 0 & 0 & \frac{1}{6}
\end{bmatrix}}.
\]
\end{example}

It is straightforward to verify that an AsTIM-$l$ matrix exists for any input vector $v$ and parameter $l$. For deterministic constructions of the nonzero submatrix, several classical choices are available, such as Hilbert and Vandermonde matrices. It is worth noting, however, that both Hilbert and Vandermonde matrices are typically ill-conditioned. While this does not affect our theoretical results, it may lead to numerical instability in practice due to rounding errors. To address this issue, one may instead construct the nonzero entries of an AsTIM-$l$ matrix by drawing them independently from a standard normal distribution $\mathcal{N}(0,1)$. With probability one, the resulting submatrix will be totally invertible.
The primary role of AsTIM-$l$ matrices in our framework is to facilitate a reliable transformation from the 1bCS measurement vector $y$ to a corresponding group testing outcome vector $y'$. This enables us to leverage efficient group testing decoding algorithms for recovering $\supp(x)$. 

We now present the construction of the sensing matrix $A$ used in the EDOCS-EE scheme in Algorithm~\ref{alg:edocsee}, and 
Algorithm~\ref{alg:randec} presents the decoding process. 

\begin{algorithm}[!]
\caption{EDOCS-EE: Sensing matrix \label{alg:edocsee}}
\begin{algorithmic}[1]
    \Require Signal vector dimension $n$, and upper bound on the support of signal vector $k$. Let $m_B:=16k\log_2 n$ and $k_0:=10\log_2 k$.
    \State Take the design matrix of fast binary splitting group testing matrix in~\cite{wang2023quickly}, denote this matrix as $B$. By Section \ref{sec:fbs}, this matrix has $m_B$ rows. 
    
    \State Partition $B$ row-wise into $[B^1;B^2;...;B^{m_B}]$.

    \State Initialize empty matrix $A$. 
    
    \For{$i=1,2,...,m_B$}
    \State Find an AsTIM-$k_0$ of $B^i$, call it $C^{(i)}$.
    \State Concatenate $C^{(i)}$ to $A$ vertically.  
    \EndFor
    \State \Return $A$
\end{algorithmic}
\end{algorithm}

\begin{algorithm}[!]
\caption{EDOCS-EE: Decoding algorithm \label{alg:randec}}
\begin{algorithmic}[1]
    \Require Notations $n$, $k$, $A$,  $B$, $m_B$, $k_0$ from Algorithm~\ref{alg:edocsee}, and the result vector $y=\nz(Ax)$.
    \State Partition $y=[y^{(1)};y^{(2)};...;y^{(m_B)}]$, where each $y^{(i)}$ is of length $k_0$. 
    
    \State Initialize zero vector $y'=0^{m_B}$. 
    
    \For{$i=1,2,...,m_B$}
    \If {$y^{(i)}\neq 0^{k_0}$} \State $y'_i=1$.
    \EndIf
    \EndFor
    \State Apply the decoding algorithm of the fast binary splitting on $B$ and $y'$, get $S$ as output. 
    \State \Return $S$
\end{algorithmic}
\end{algorithm}

We state the main theorem of this section in Theorem~\ref{thm:ee} below: 
\begin{theorem}~\label{thm:ee}
    \emph{(EDOCS-EE)} Let $k_0:=10\log_2 k$, 
    there exists a one-bit compressed sensing matrix $A\in\mathbb{R}^{m\times n}$ for exact support recovery of a $k$-sparse signal vector whose support is drawn uniformly at random, with $m=16k_0 \cdot k\log_2 n=O( k\log k\log n)$ measurements (Algorithm~\ref{alg:edocsee}), and a recovery algorithm (Algorithm~\ref{alg:randec}) with runtime $D=O(m)$ , that fails with probability $<\epsilon_1+\epsilon_2$, where $\epsilon_1=20\log_2(n)\cdot k^{-9}$ and $\epsilon_2=e^{-k}+n^{-k}+5k^{-3}$.
\end{theorem}
\begin{proof}
   We construct the sensing matrix $A$ using Algorithm~\ref{alg:edocsee} and perform decoding via Algorithm~\ref{alg:randec}. Throughout, we adopt the notation from these algorithms unless otherwise stated.

\emph{Correctness.}
By the connection between group testing and the $1$-bit compressed sensing problem discussed in Section~\ref{sec:gt1bcs}, it suffices to show that the vector $y'$ in Algorithm~\ref{alg:randec} is reliable. Specifically, we must rule out the occurrence of \emph{accidental zeros}, i.e., errors introduced when translating the $1$-bit CS measurements $y$ into the group testing outcomes $y'$.

An accidental zero occurs if there exists $i \in [m_B]$ such that
\begin{equation}\label{eqn4}
    y^{(i)} = \nz(C^{(i)}x) = 0^{k_0}
    \text{ while }
    \supp(B^{i}) \cap \supp(x) \neq \varnothing.
\end{equation}

We show that this event occurs with vanishing probability. By Lemma~\ref{lem:error2}, with high probability, no test in $B$ contains more than $k_0$ defectives, so
\[
|\supp(B^{i}) \cap \supp(x)| \leq k_0
\text{ for all } i \in [m_B].
\]

Conditioned on this event, the matrix $C^{(i)}$ (which is AsTIM-$k_0$) guarantees that $C^{(i)}x \neq 0^{k_0}$ whenever $\supp(B^{i}) \cap \supp(x) \neq \varnothing$. Therefore, the event in~\eqref{eqn4} cannot occur.

\emph{Complexity and error probability.}
The measurement and decoding complexities follow directly from the construction and are therefore omitted (Recall Lemma~\ref{lem:error1}, claiming that decoding complexity of the fast binary splitting algorithm is asymptotically the same as its number of tests $O(k\log n)$.) The overall error probability is given by combining Lemma~\ref{lem:error1} and Lemma~\ref{lem:error2}.
\end{proof}

\section{Conclusion}\label{sec:sum}
In this paper, we proposed two schemes for support recovery in the one-bit compressed sensing (1bCS) problem under different recovery guarantees. 

For universal support recovery, we developed an $\epsilon$-approximate scheme with
$m=O(k\epsilon^{-1}$ $\log n\log(n/k))$ measurements and decoding complexity $O(\epsilon^{-1}m)$, and an exact recovery scheme with $m=O(k^2\log n\log (n/k))$ measurements and decoding complexity $O(km)$. Compared to the state-of-the-art results in~\cite{acharya2017improved,matsumoto2023improved}, our schemes incur an additional $O(\log n)$ factor in the number of measurements, but achieve significantly improved decoding efficiency. In particular, the decoding complexity is reduced from $O(n\epsilon^{-1}\log(n/k))$ to $O(k\epsilon^{-2}\log n$ $\log(n/k))$ for $\epsilon$-approximate recovery, and from $O(nk\log n)$ to $O(k^3\log n\log(n/k))$ for exact recovery. To the best of our knowledge, these are the first universal 1bCS support recovery schemes achieving sublinear decoding complexity.

For probabilistic exact support recovery, we improved upon the results of~\cite{yang2025sublinear} by reducing the number of measurements from $O(k^2\log^2 n(\log^2 k+(\log\log n)^2))$ to $O\!\left(k\log k\log n\right)$, while also achieving faster decoding.

Several directions remain for future work. A natural goal is to design schemes that simultaneously achieve optimal measurement complexity and sublinear decoding time. This appears challenging due to the presence of the \emph{accidental zero} issue (see Section~\ref{sec:gt1bcs}), which has no direct analogue in group testing. Another direction is to develop universal $\epsilon$-superset recovery schemes~\cite{matsumoto2023improved} with sublinear runtime that outperform our exact recovery construction. Extending our framework to noisy settings is also of significant interest. For instance, one may consider the bounded Massart noise model~\cite{awasthi2016learning}, where each measurement outcome is independently flipped with probability $\eta < 1/2$. 

More broadly, an intriguing research direction is to better understand the role of computationally efficient decoding in information-theoretically optimal  compression schemes, and whether the ideas developed here for one-bit compressed sensing can inspire analogous encoder--decoder designs in other data compression settings. We hope that this work contributes toward a broader perspective in which decoding complexity, alongside optimality of compression, is regarded as a fundamental consideration in the design of practical compression systems.

\section*{Acknowledgment}
The authors would like to thank 
anonymous reviewers of a previous version of this paper for pointing out a technical issue that we have since corrected, and also for their suggestions for improvement. 

\bibliographystyle{alpha}
\bibliography{references}

\end{document}